\documentclass[]{spie}  

\usepackage{amsmath,amsfonts,amssymb}
\usepackage{graphicx}
\usepackage[colorlinks=true, allcolors=blue]{hyperref}

\title{Design of the new CHARA instrument SILMARIL: pushing the sensitivity of a 3-beam combiner in the H- and K-bands}

\author[a]{Cyprien Lanthermann}
\author[a]{Theo ten Brummelaar}
\author[b]{Peter Tuthill}
\author[c]{Marc-Antoine Martinod}
\author[a]{E. Robert Ligon}
\author[d]{Douglas Gies}
\author[a]{Gail Schaefer}
\author[a]{Matthew Anderson}
\affil[a]{The CHARA Array of Georgia State University,         Mount Wilson Observatory, Mount Wilson, CA 91023, USA}
\affil[b]{School of Physics, Sydney University, N.S.W. 2006, Australia}
\affil[c]{Institute of Astronomy, KU Leuven, Celestijnenlaan 200D, 3001, Leuven, Belgium}
\affil[d]{Center for High Angular Resolution Astronomy and Department of Physics and Astronomy, Georgia State University, P.O. Box 5060, Atlanta, GA 30302-5060, USA}

\authorinfo{Further author information: (Send correspondence to C.L.)\\C.L.: E-mail: clanthermann@gsu.edu}

\begin{document} 
\maketitle

\begin{abstract}
Optical interferometry is a powerful technique to achieve high angular resolution. However, its main issue is its lack of sensitivity, compared to other observation techniques. Efforts have been made in the previous decade to improve the sensitivity of optical interferometry, with instruments such as PIONIER and GRAVITY at VLTI, or MIRC-X and MYSTIC at CHARA. While those instruments pushed on sensitivity, their design focus was not the sensitivity but relative astrometric accuracy, imaging capability, or spectral resolution. Our goal is to build an instrument specifically designed to optimize for sensitivity. This meant focusing our design efforts on different parts of the instrument and investigating new technologies and techniques. First, we make use of the low noise C-RED One camera using e-APD technology and provided by First Light Imaging, already used in the improvement of sensitivity in recent new instruments. We forego the use of single-mode fibers but still favor an image plane design that offers more sensitivity than a pupil plane layout. We also use a minimum number of optical elements to maximize the throughput of the design, using a long focal length cylindrical mirror. We chose to limit our design to 3 beams, to have the capability to obtain closure phases, but not dilute the incoming flux in more beam combinations. We also use in our design an edge filter to have the capability to observe H- and K-band at the same time. We use a low spectral resolution, allowing for group delay fringe tracking but maximizing the SNR of the fringes for each spectral channel. All these elements will lead to a typical limiting magnitude between 10 and 11 in both H- and K-bands.
\end{abstract}


\section{INTRODUCTION}
\label{sec:intro}  

Interferometer arrays possess between one to two orders of magnitude gain in spatial resolution over the largest single-aperture telescopes currently under development. The CHARA Array’s longest baseline yields angular diameter measurements as small as 0.7 mas in K-band improving to 0.2 mas in R-band. CHARA can resolve stellar disks all along the main sequence from O- to M-types, whereas there will be no main sequence stars resolvable by the next generation of single aperture telescopes. 

A boost in sensitivity will expand all of our observing programs, and in particular those involving the faintest targets like AGN cores, YSO’s, binary stars in different stages of evolution, faint dwarf stars, and exoplanet hosts.

The CHARA Array’s six 1-m aperture telescopes are arranged in a Y-shaped configuration yielding 15 interferometric baselines from 34 to 331 meters and 10 independent closure phases. These are the longest OIR baselines yet implemented and permit resolutions at the sub-milliarcsecond (mas) level. The facility’s primary components and sub-systems are described more fully by ten Brummelaar et al.\ (2005)~\cite{2005ApJ...628..453T}

CHARA Classic (ten Brummelaar et al.\ 2005~\cite{2005ApJ...628..453T}) is a two-beam, J-, H- and K-band open-air, beamsplitter-based system providing visibility amplitude measurements optimized for sensitivity in the NIR. We routinely observe science targets as faint as 7.5 magnitude in H- and K-bands, and in good seeing we reach magnitude 8.

CLIMB (ten Brummelaar et al.\ 2013~\cite{2013JAI.....240004T}) is an expansion of CLASSIC able to combine three beams and obtain three visibility amplitudes and one closure phase. Since there are six beams in the Array, there are two independent sets of CLIMB optics, the second of which can be reconfigured as CLASSIC. The magnitude limit of CLIMB is not as good as CLASSIC as the beams must be divided three ways instead of two resulting in a loss of signal to noise ratio (SNR).

However, two new beam combiners have recently been implemented at the CHARA Array, MIRC-X\cite{MIRCX} and MYSTIC\cite{MYSTIC,SetterholmSPIE2022}, reaching the same limit in magnitude as CLASSIC/CLIMB, but recombining all six telescopes. Therefore, the usefulness of CLASSIC/CLIMB has decreased. 

The most significant challenge in ground based interferometry is sensitivity, so we propose to increase the faint limit of the CHARA Array’s CLASSIC/CLIMB beam combiner by 2 magnitudes in the near infrared. This will be achieved by replacing the 20-year-old PICNIC detector with a modern SELEX MOVPE SAPHIRA\cite{gert2016,2018arXiv180508419G} based detection system. This upgrade will also include spectral resolution in both H- and K-bands, and simultaneous observations in both bands.

We describe the different elements allowing SILMARIL to push for sensitivity in Section~\ref{sec:pushsens}. We then present the theoretical performance that SILMARIL should achieve in
Section~\ref{sec:simulations}.  Software issues are 
considered in Section~\ref{sec:soft}.
Increasing the faint limit opens up a larger volume of parameter space for investigation and offers more opportunities to study rare objects not found in the solar neighborhood. A few of these new scientific opportunities are described in Section~\ref{sec:sci}. We offer conclusions in Section~\ref{sec:concl}.

\section{Push for sensitivity}\label{sec:pushsens}

The optical design of SILMARIL is focused on maximizing sensitivity, while making as few compromises as possible in the other aspects of the instrument. 

\subsection{e-APD Detector}

The main element that will improve sensitivity is the use of the new SAPHIRA detector\cite{gert2016,2018arXiv180508419G}, using  e-APD technology. This detector was built specifically for use in the Leonardo Adaptive Optics (AO) and Fringe Tracker (FT) of the GRAVITY instrument at VLTI~\cite{GRAVITY}. This detector has the advantage of having a sub-electron readout noise, with a fast frame rate, and a relatively decent dark current. All these characteristics are what is needed for interferometric instruments as well as AO systems.
The SAPHIRA detector is now used by several new interferometric instruments, including the MIRC-X~\cite{MIRCX,2019A&A...625A..38L} and MYSTIC~\cite{MYSTIC,SetterholmSPIE2022} instruments at the CHARA Array that have a ready-to-use camera, the C-RED One~\cite{CRED}, employing the SAPHIRA detector and built by First Light Imaging.

Because those cameras are already successful at the CHARA Array and we have the technical expertise to use them, we also chose to adopt this ready-to-use solution, which will make it easier to develop the necessary software, as explained in Section~\ref{sec:soft}.

\subsection{3-Telescope Beam Combiner}

The effect of the number of beams that an interferometric instrument will recombine is still debated in the community. Some claim that the SNR is not impacted by the number of beams recombined, others claim that the more beams you recombine, the lower the SNR. From our experience with CLASSIC (2-Telescope beam combiner) and CLIMB (3-Telescope beam combiner), using the same optical design and camera, we can notice that CLIMB has a lower sensitivity than CLASSIC. We conclude that, at least in this case, the more beams you combine, the less sensitivity you attain. 

Thus, for SILMARIL, we are conservative and limit the number of telescopes we combine to three in order to limit the loss in sensitivity, while still being able to measure one closure phase.

\subsection{Image Plane Combiner}

The original plan for the improvement of the sensitivity of CLASSIC/CLIMB was to reuse the original design, limiting the change of optics, and installing the new C-RED One camera. However, as we want to push as far as possible for sensitivity, we ran simulations to compare the performances of a 3-beam pupil plane design, such as CLIMB, and a 3-beam image plane design. 

The parameters adopted in the simulations are summarized in Table~\ref{tab:planesimparam}. The details of these simulations are the subject of another paper (Tuthill et al. in prep.).

\begin{table}[ht]
\label{tab:planesimparam}
\begin{center}       
\caption{Model Beam Combiner Parameters} 
\begin{tabular}{|c|c|c|} 
\hline
\rule[-1ex]{0pt}{3.5ex}  Parameter & Pupil Plane Design & Image Plane Design  \\
\hline
\rule[-1ex]{0pt}{3.5ex}  Readout Rate (Hz) & 750 & 83.3   \\
\hline
\rule[-1ex]{0pt}{3.5ex}  Sample Time (ms) & 1.33 & 12   \\
\hline
\rule[-1ex]{0pt}{3.5ex}  Photometry per pixel (N$_{ph}$) & 1  & 0.04   \\
\hline
\rule[-1ex]{0pt}{3.5ex} Delay Range ($\mu$m) & $\pm$ 50 & $\pm$ 50 \\
\hline
\rule[-1ex]{0pt}{3.5ex}  Spectral Dispersion R & 1 & 30\\
\hline
\rule[-1ex]{0pt}{3.5ex}  Central Wavelength ($\mu$m) & 1.673  &  1.673 \\
\hline
\rule[-1ex]{0pt}{3.5ex}  Bandwidth Per Channel ($\mu$m) & 0.285  & 0.056 \\
\hline
\rule[-1ex]{0pt}{3.5ex}  Number of Spectral Channels & 1 & 5  \\
\hline
\rule[-1ex]{0pt}{3.5ex}  Number pixels across fringes & 3 -- 6 -- 9 & 5 -- 10 -- 15\\
\hline
\rule[-1ex]{0pt}{3.5ex}  Pixels read per sample & 1 & 225 \\
\hline 
\rule[-1ex]{0pt}{3.5ex}  Field of View & 3.78 $\mu$rad & Airy Disk   \\
\hline
\rule[-1ex]{0pt}{3.5ex}  Integration Time (s) & 0.72 & 0.72  \\
\hline
\rule[-1ex]{0pt}{3.5ex}  Samples per integration & 538 & 60  \\
\hline
\rule[-1ex]{0pt}{3.5ex}  Total pixels per integration & 538  & 13500  \\
\hline
\rule[-1ex]{0pt}{3.5ex}  Camera Readout Noise (e$^-$ ) & 0.7 & 0.7 \\
\hline
\rule[-1ex]{0pt}{3.5ex}  Camera Dark Current (e$^-$p$^{-1}$s$^{-1}$) & 80  & 80 \\
\hline
\rule[-1ex]{0pt}{3.5ex}  Camera Excess Noise Factor & 1.25  & 1.25 \\
\hline
\end{tabular}
\end{center}
\end{table} 

For each simulations, we proceed in four steps:
\begin{itemize}
    \item The creation of a phase mask, representing the atmospheric turbulence
    \item The simulation of the instrument and the corresponding interferometric signal 
    \item The addition of the detector noise on the signal
    \item The measurement of the SNR from the extraction of the fringe peak
\end{itemize}

\begin{figure}[h]
\makebox[\textwidth]{\includegraphics[width=0.7\textwidth]{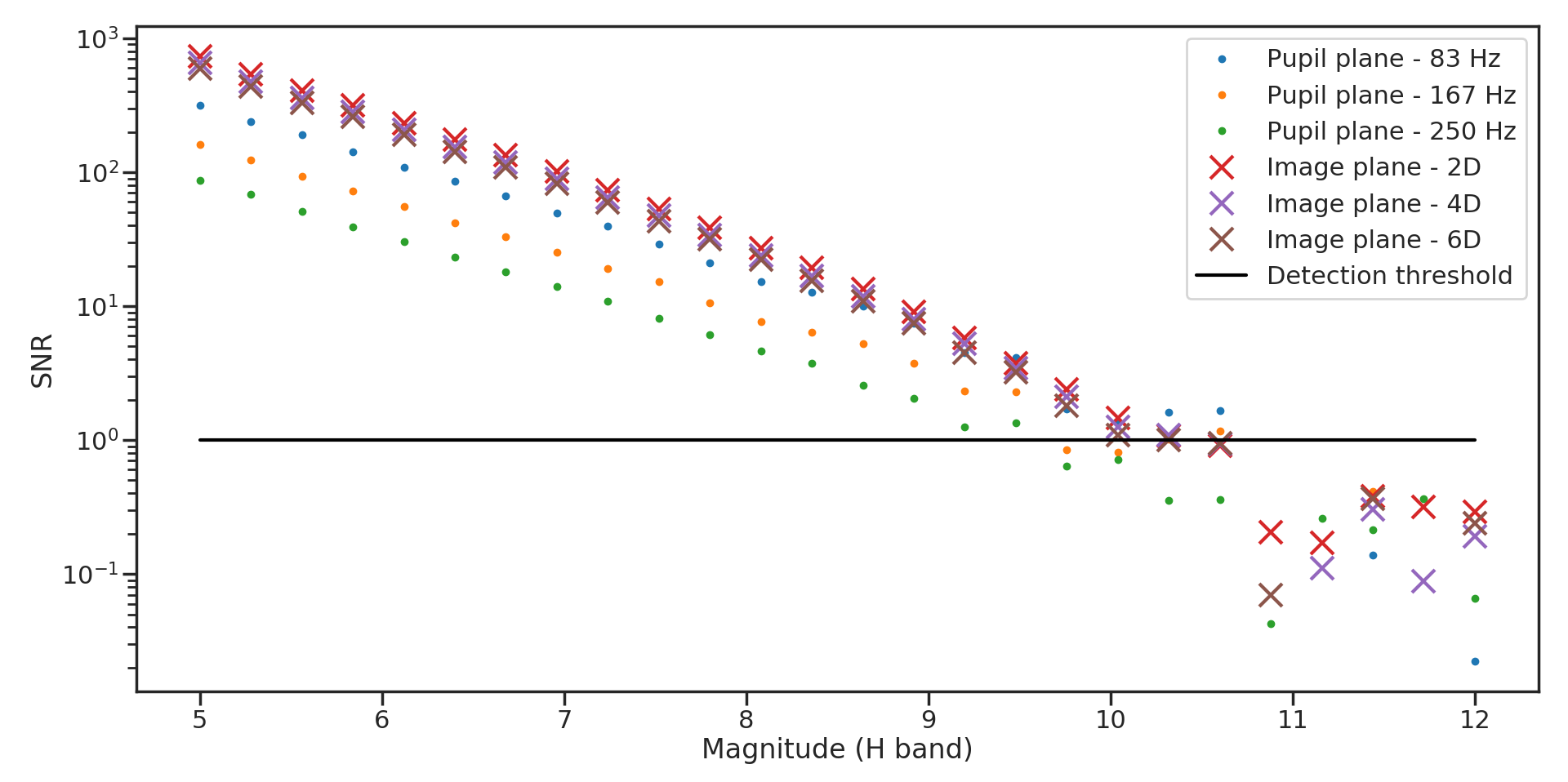}}
\caption{Signal to Noise ratio for different parameters of Pupil and Image plane beam combiners
}\label{fig:ressimplane}
\end{figure}

The results of the simulations are presented in Figure~\ref{fig:ressimplane}.
In this figure, we display the SNR as a function of the magnitude of the target, in the H-band, for different parameters, corresponding to the 3 different combination of 2 of the 3 telescopes. Image plane, D represents the diameter of the beam (19 mm), xD representing then the distance separating the 2 combined beams expressed in number of beams.

We can see that the SNR of the image plane simulation is always better than that of the pupil plane. With an image plane design we should be able to gain an extra 1 magnitude compared to a pupil plane design.

We therefore chose to switch from the initial pupil plane design of CLIMB to a new image plane design described in Section~\ref{sec:minidesign}.

\subsection{Minimal Design}\label{sec:minidesign}

\begin{figure}[h]
\makebox[\textwidth]{\includegraphics[width=\textwidth]{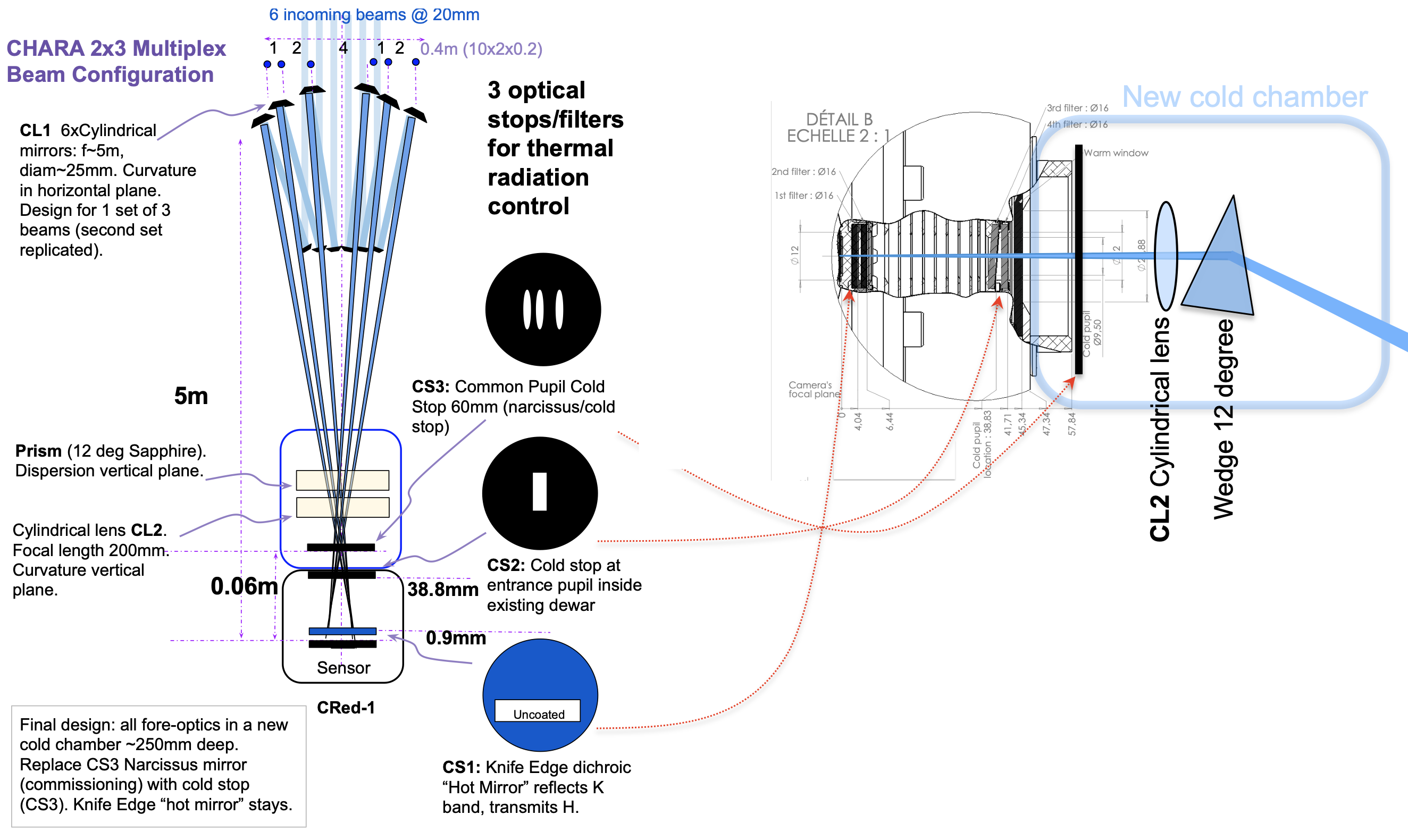}}
\caption{Scheme of the optical design of SILMARIL.
}\label{fig:SILMARILdesign}
\end{figure}

The optical design of SILMARIL is inspired from the FOURIER instrument~\cite{2020SPIE11446E..0VM} at MROI. The concept is to keep the number of optics used to a minimum to avoid the loss of flux that each optical element introduces.
A scheme of the design is shown in Figure~\ref{fig:SILMARILdesign}.

The first important element is the use of very long focal length (5.41m) cylindrical (CL1) mirrors to form the image in the fringe axis (horizontal) and a much shorter focal length (0.35m) second cylindrical lens (CL2) to form the image in the spectrally dispersed axis (vertical). The intent, based on the desire to optimize for faint light operation, is to minimize the number of optics and reduce background in K-band as much as possible, while at the same time allowing a development path that enables us to test the fundamental optical design on the sky without using a second dewar. For ultimate performance an upgrade including a second dewar with specialized cold stops is planned.

Mirrors at the standard beam height are used to send the beams to long focal length cylindrical mirrors CL1 at a higher level. These cylindrical mirrors are arrayed in the standard 2-4-6 non-redundant beam pattern, and they send the beams to a common focal point in the horizontal axis. These are followed by a second cylindrical lens CL2 acting in the vertical axis and of shorter focal length. This compresses the image to a small number of pixels in the vertical axis. When used in concert with a vertical dispersive element, the instrument now yields a modest spectral resolution (R$\sim$30), a layout usually termed channel spectrum fringes. This low spectral resolution allows for group delay tracking of the fringes.  An extra fold mirror is necessary to fit things into the available space.

This arrangement has minimal optics and will produce an image that can be placed directly onto the C-RED One camera. At a later time, we intend to add a fore-dewar based largely on the MYSTIC~\cite{MYSTIC,SetterholmSPIE2022} design.

One primary area of concern is contamination in the H-band from background counts in the K-band. This can be resolved by replacing the first filter in the camera (closest to the detector) with a custom "edge filter" that has an H-band coating only on the lower half. There are two problems with an edge filter such as this, both to do with how much K-band light reaches the H-band part of the detector.

The first issue is that some of the background light will be diffracted by the edge of the filter and reach the longest wavelength channel of the H-band system. The Fresnel scale for diffraction is given by
\begin{align}
    d_F = \sqrt{\frac{x  \lambda}{\pi}}
\end{align}
where $x$ is the distance from the refracting edge and the first filter in the camera is 2.4 mm away from the detector. For K-band we will have a Fresnel scale of 40 $\mu$m. For the 24 $\mu$m pixel size and a gap of 6 pixels on the chip between H and K, we have 3.2 Fresnel scale units, resulting in much less than 1\% of the K-band background light diffracting into the longest wavelength channel of the H-band system and much less in the other pixels.

\begin{figure}[h]
\makebox[\textwidth]{\includegraphics[width=0.8\textwidth]{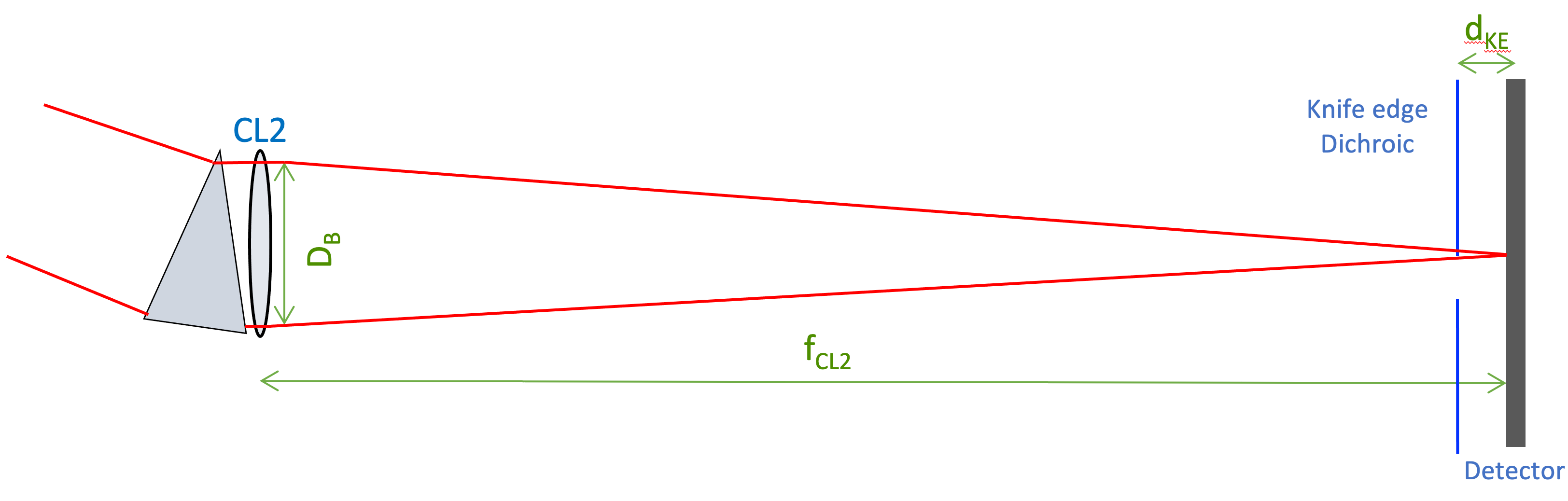}}
\caption{Geometry showing contamination of H-band by K-band light
}\label{fig:KinH}
\end{figure}

A more serious concern is the geometric contamination of light from K-band in the converging beam as shown in Figure~\ref{fig:KinH}. 
The K-band light will no longer reach the chip after a distance of
\begin{align}
    \frac{d_{KE}}{f_{CL2}} \times D_B.
\end{align}
For the focal length of 350mm and the filter to detector distance $d_{KE}=2.4$mm, this will be 130$\mu$m or about 5.5 pixels while the gap between H- and K-bands is 6 pixels. This could be improved by either increasing the focal length of the CL2 optic, thereby reducing the spectral resolution of the system, or placing the filter surface much closer to the detector surface.

\begin{figure}[h]
\makebox[\textwidth]{\includegraphics[width=0.25\textwidth]{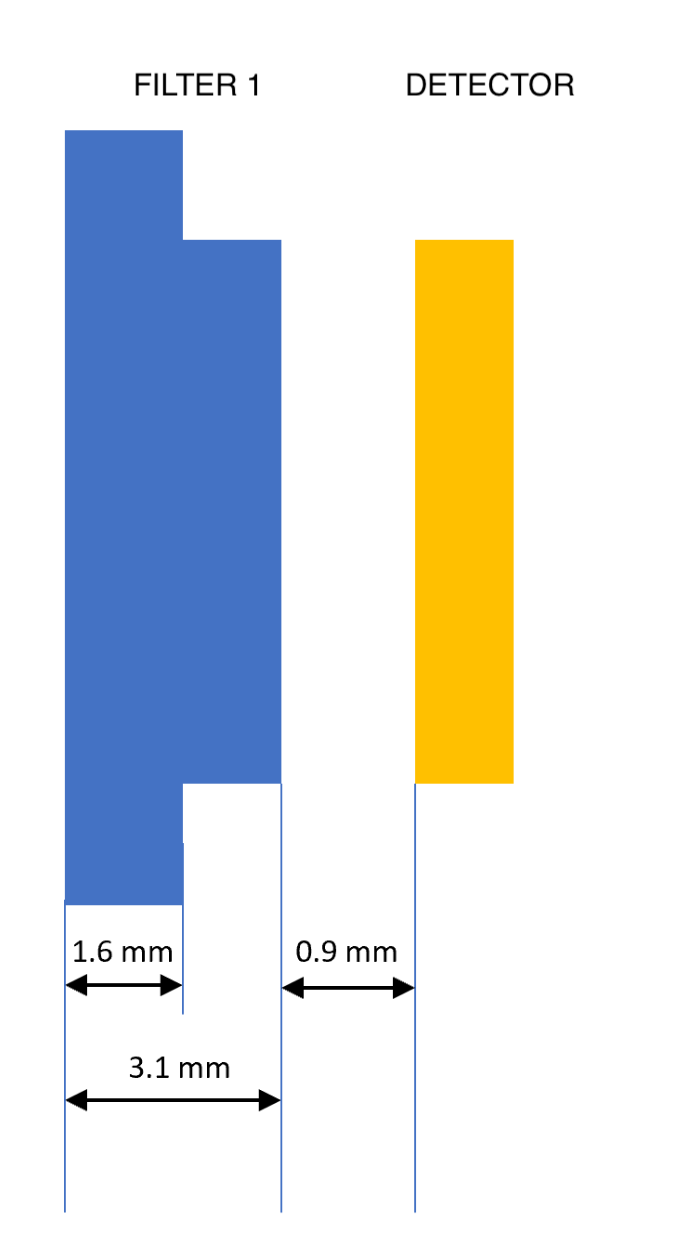}\includegraphics[width=0.65\textwidth]{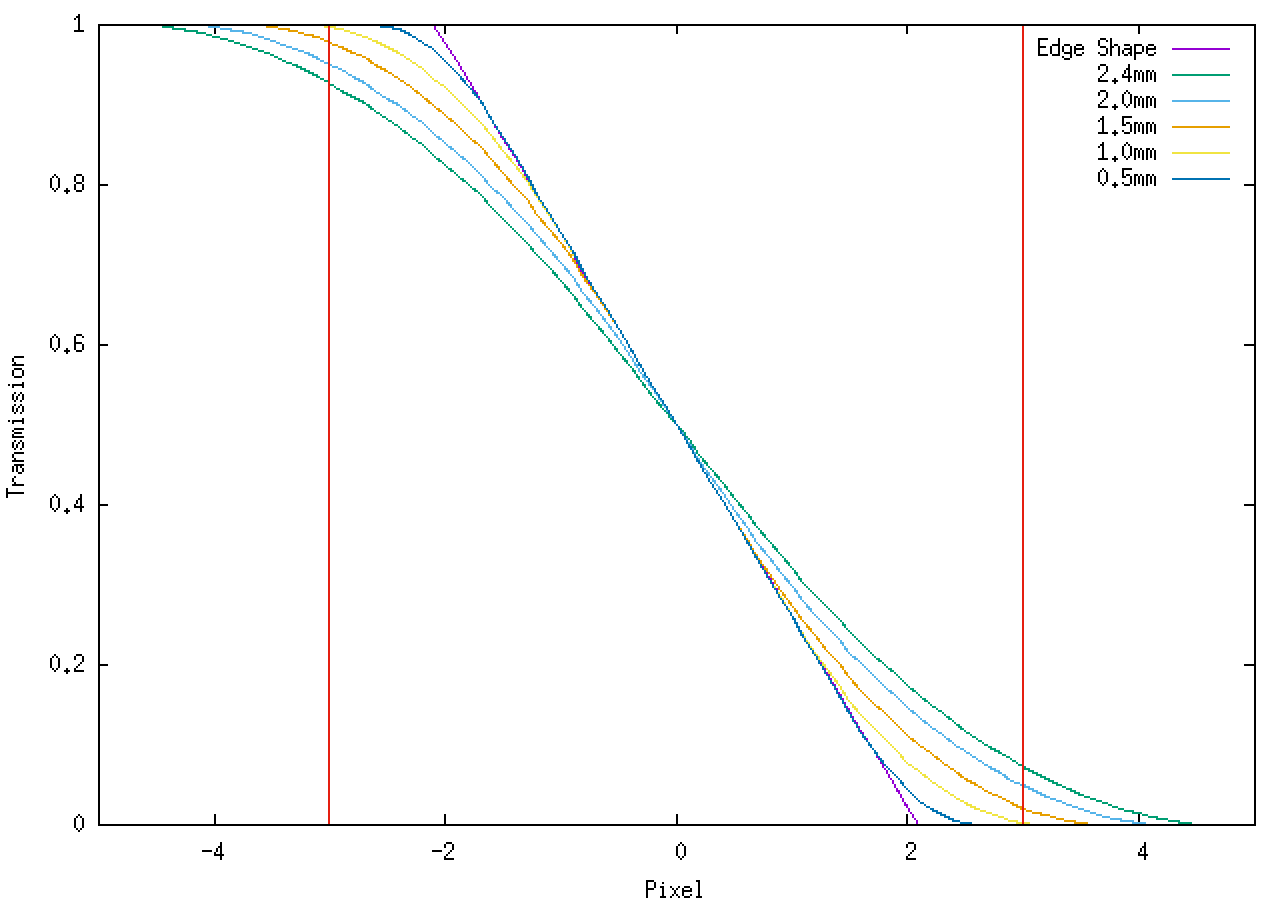}}
\caption{Left: Proposed shape of edge filter. Right: Effect of $d_{KE}$ on the geometry of the contaminated area, with a transition width of 100$\mu$m.
}\label{fig:edgefiltscheme}
\end{figure}

In Figure~\ref{fig:edgefiltscheme} on the left, we show a filter substrate design that will allow us to get to within 0.9mm of the chip. On the right we show the effect of the distance $d_{KE}$ on the geometry of the contaminated area. A design like this would reduce the area of contamination to 2.5 pixels. However, getting a filter with such a design and getting closer to the detector will increase the risk of damaging the detector if an incident should happen with the filter. Furthermore, this calculation assumes a perfectly sharp edge on the filter that turns out to be very difficult to make. After numerous discussions with optical vendors, we have found that this custom shape filter and sharp edge boundary (specified as $<$50$\mu$m) is very difficult to acquire. We contacted 10 companies who we have either used before or have been recommended by First Light or other members of the consortium. Most put in a no-bid, while the two that did bid gave prices of \$18k and \$50k. A third vendor has told us that they cannot make the odd shape we are asking for, as it will not fit inside their coating chamber, but they can make a filter of the standard shape for the camera with a transition zone of the filter edge as small as 100$\mu$m for a price of \$ 5k. However, after studying more in details all the requirements of the edge filter, it appeared that this company couldn't meet them and decided not to bid anymore.

The amount of K-band light getting through now becomes an integral of the edge shape, assumed to be linear, and the beam shape in this dimension is rectangular. We show in Figure~\ref{fig:edgefiltscheme}, right, the final transition shape for a 100$\mu$m edge boundary at various distances from the detector. At the default distance of 2.4mm we have less than 10\% contamination of background light in the longest wavelength channel of the H-band. We conclude that using the standard design, rather than a higher risk specialized shape, and a specification of 100$\mu$m on the boundary edge will be fine for our needs.

The final consideration is the shape of the cold input pupil. The beams are anamorphic so rather than use the default F-20 circularly symmetric cold stop, we plan to use a rectangular or square shape. First Light have done a preliminary design of a rectangular cold stop.

\begin{figure}[h]
\makebox[\textwidth]{\includegraphics[width=0.8\textwidth]{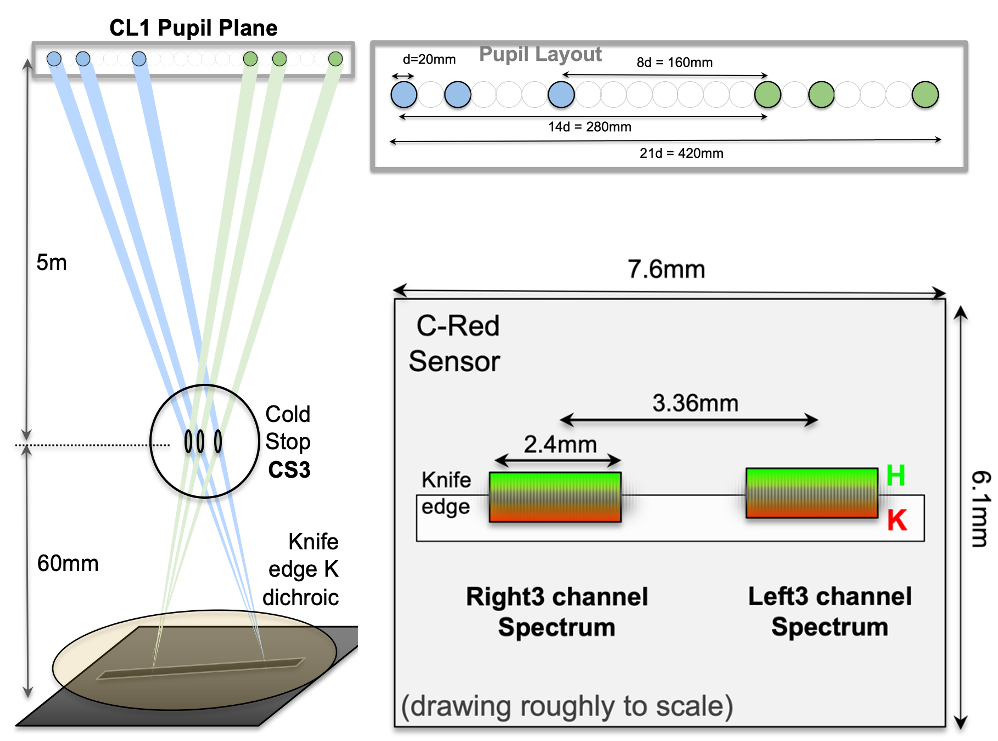}}
\caption{Left: Beam layout for 2$\times$3 way beam combination. Above right: Geometry of beam layout. Below right: Position of fringes on detector. Note that the dimensions in this diagram are nominal and based on estimates made at the time it was drawn.
}\label{fig:finallayout}
\end{figure}

We plan in the end to implement a second set of 3 beams that would fit in the current design, to obtain a 2$\times$3-telescope beam combiner, making use of the six telescopes of the CHARA Array at the same time. A scheme of the final layout is presented in Figure~\ref{fig:finallayout}.

\section{Performance Simulations} \label{sec:simulations}

With a design with parameters fixed, we can perform a simulation of the performance of such a design. The SILMARIL performance simulation consists of two programs. 

The first one simulates the effect of the atmospheric turbulence on the fringe phase seen by the detector. This model is based on the one described in CHARA Technical Report~\#103\footnote{\href{https://www.chara.gsu.edu/astronomers/technical-reports}{https://www.chara.gsu.edu/astronomers/technical-reports}}. The parameters we changed for this first part of the simulation are:

\begin{itemize}
    \item The $r_0$ parameter (r): this seeing parameter is a measure of the stability of the atmospheric turbulence. The higher it is, the better the seeing, and the more stable the fringes will be. In this study we used 3 different values for this parameter: $r_0 = 5$~cm, representing bad seeing, $r_0 = 10$~cm representing an average seeing and $r_0 = 20$~cm, representing great seeing.
    
    \item The spectral resolution (R): this parameter will determine the number of pixels in the spectral dimension. The advantage of having a lower spectral resolution is to disperse the light over fewer pixels, giving a better signal to noise per pixel. The advantage of having a higher spectral resolution is having a bigger gap between the H and K band, making the transition of transmission of the edge filter less critical. Another advantage is to have a longer coherence length, which can make it easier to track the fringes. For this study, we fixed this parameter to R = 35, which is the initial value we have assumed for this parameter.
    
    \item The number of pixels to sample the fringe dimension (n). This number depends on the number of pixels we want to use to sample one fringe and the spectral channel we sample, as the size of the fringe depends on the wavelength we probe. As we fixed the sampling of the fringes to 2.5 pixel/fringe, the parameter n goes from 43 to 50 pixels in the H-band and from 54 to 64 pixels in the K-band (for the previous K filter; the current K’ filter gives from 57 to 65 pixels). For the H-band we used n = 43, 47, and 50. For the K-band, we used n = 54, 59, and 64.
    
    \item The spectral bandwidth parameter (l): this parameter allows us to specify the spectral band we want to work on, by specifying the central wavelength and the bandwidth. For this study, we used the default H-band and the previous specification of the K-band. This parameter will play a role in the number of pixels over which the light will be dispersed, along with the spectral resolution parameter R. It will also play a role in the atmospheric turbulence, as its parameters change as a function of the wavelength. The larger the wavelength, the more stable is the atmospheric turbulence.
\end{itemize}

The second program simulates the detector itself, including sample time and incoherent integration time, simulating its performance and the tracking of the fringes. In addition, the script computes the SNR of the fringes. To compute the SNR for the signal, we take the maximum of the power spectrum in the position of a fringe (3 pixels large), for each pair of telescopes. For the noise, we compute the RMS of all the pixels in the power spectrum that are not in the area of a fringe, while not including the low frequencies where there is a great deal of power due to the atmospheric noise.

The parameters we changed for this second part of the simulation are:
\begin{itemize}
    \item The magnitude of the observed star (c): this parameter allows us to probe the limiting magnitude we can observe with SILMARIL. The model used for the expected number of photons is the same as that was used for TR~\#103 and the original proposal. For this parameter, we use c = 7.0, 8.0, 9.0, 10.0, and 11.0.
    
    \item The number of frames we incoherently add (s): this is the same as changing the sample time of the detector and can increase the SNR of the fringes, increasing the sensitivity of the instrument. The first program does one calculation of the atmosphere for every millisecond. Adding these together allows us to simulate longer sample times. However, if we average too much, the fringes will start to get blurred, and we will start to decrease the SNR of the fringes. The better the seeing, the longer we should be able to add frames incoherently before we start to blur the fringes. For this parameter, we used: s = 1, 3, 6, 9, 12, and 15.
    
    \item The number of power spectra we coherently add (p). As for the sample time, increasing the number of power spectra frames we add can increase the SNR. But if we add too many, then the atmospheric noise will overwhelm the signal and reduce the SNR.
    
    \item The parameters of the camera, D: these parameters include the readout noise (RON), the excess noise factor (ENF), and the background (BKG). For these simulations, we fixed the RON to 1.0, which is a conservative approximation for the RON of the C-RED One camera, and the ENF to 1.4, which is the characterization of the C-RED One camera’s ENF. For the background, we used BKG = [41, 129, 408, 4082], which correspond, respectively, to the background with the C-RED One with a MYSTIC type second Dewar, without the second Dewar but with a geometric mean of NIRO type improvement and C-RED One, just the NIRO type improvement, and just the C-RED One.
\end{itemize} 

We ran simulations for all of the combinations of the different values of parameters discussed previously. We then plot the SNR as a function of the stellar magnitude for each set of different parameters. To compute the SNR, we take the mean of SNR over the frames for each group of telescopes and then we take the mean of the 3 sets to compute the global SNR in the data. On the plot, we show the position of SNR = 2 by a line, which is the limit for which we should be able to track the fringes while observing. If the SNR is above this line, we should be able to observe the target with the set of parameters we used.

Here we describe the results for runs of the simulator for different parameters.

\subsection{Effect of the number of frames we add incoherently (s)}

The number of frames we add incoherently allows us to improve the SNR of the fringes by integrating more flux. However, if we add too many frames, the fringes will start to move on the detector (as a function of the seeing), and therefore, we would start to blur the fringes, decreasing the signal of the fringes we want to measure.

\begin{figure}[h!]
\makebox[\textwidth]{\includegraphics[width=0.5\textwidth]{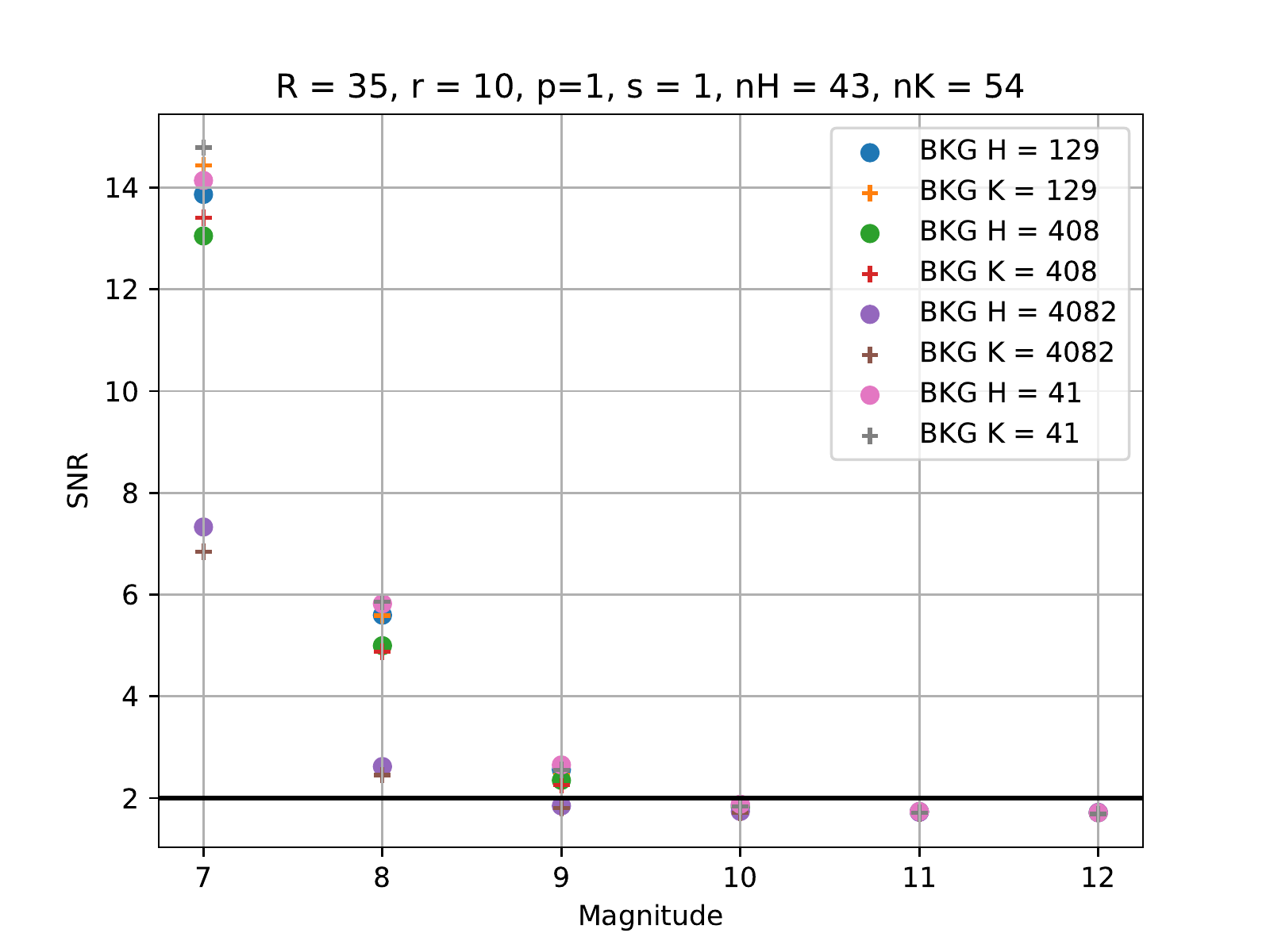}
\includegraphics[width=0.5\textwidth]{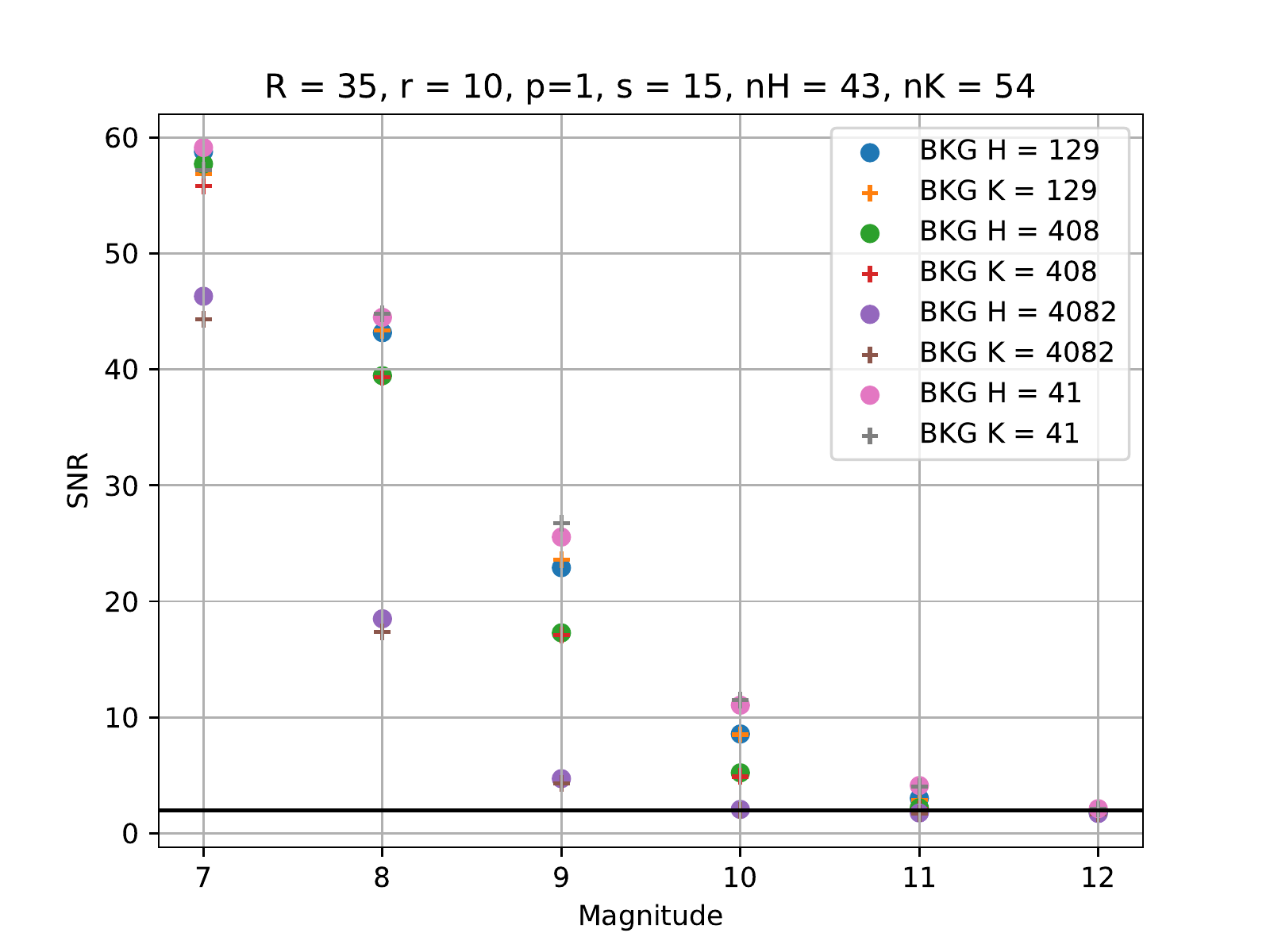}}%
\caption{SNR as a function of the star magnitude for different values of incoherent adding of frames. The parameters used on these plots are $r_0$ = 10 cm, R = 35, n = 43, p = 1. The left plot is for s = 1, the right is s = 15 frames added incoherently.}\label{fig:s}
\vspace*{0.5cm}
\end{figure}

In Fig.~\ref{fig:s}, we display the SNR as a function of the magnitude. The dots are for the H-band simulations, the crosses are for the K-band simulations. The different colors represent different levels of background with the values in legends being in e/s/pixel. The black line is showing the SNR=2, the limit for which we should be able to track the fringes (phase tracking). The different plots are for different numbers of frames incoherently added (s = 1 and 15 frames). Note that the trend on the plot is similar for the other values of s not shown here.

As expected, Fig.~\ref{fig:s} shows that increasing the number of frames we add incoherently has a big impact on the SNR, increasing the limit observable magnitude from 9 to 12.

\begin{figure}[h!]
\makebox[\textwidth]{\includegraphics[width=0.7\textwidth]{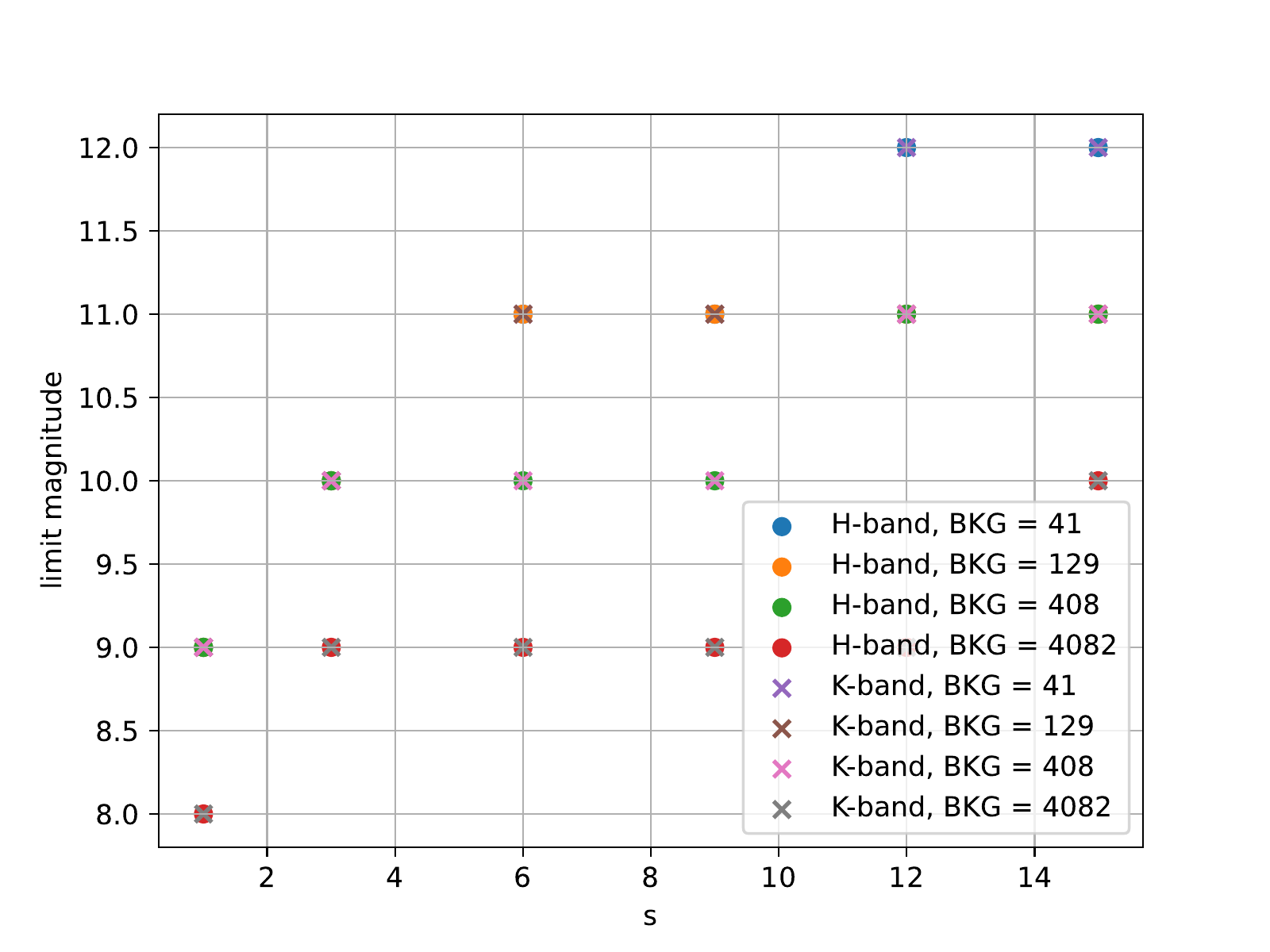}}
\caption{Limiting magnitude as a function of s, for different background levels (colors) and the different spectral bands (dots: H-band, crosses: K-band)}\label{fig:s_mag}
\vspace*{0.5cm}
\end{figure}

In Fig.~\ref{fig:s_mag}, we show the limiting magnitude as a function of number of frames s, for different background levels and the two different spectral bands.
We can first see that the limiting magnitude is the same for both spectral bands for each different set of parameters. We also notice that the best limiting magnitude increases with s, as expected. For the highest background, we need to adopt s = 15 to reach a limiting magnitude of 10, otherwise, the limiting magnitude is 9, and even 8.0 for s = 1.
For the intermediate values of background, we reach a limiting magnitude of 11 from s = 6 for BKG = 129e/s/frame and from s = 12 for BKG = 129e/s/frame. For the lowest background, we read a limiting magnitude of 12 for s = 12.

\subsection{Effect of the number of frames we add coherently (p)}

The number of power spectra we add coherently should allow us to improve the SNR of the fringes by integrating more signal in the power spectrum. However, if we add too many power spectra, the signal of the fringes will start to move on the detector (as a function of the seeing), and therefore, we would start to blur their signal in the resulting power spectrum.

\begin{figure}[h!]
\makebox[\textwidth]{\includegraphics[width=0.5\textwidth]{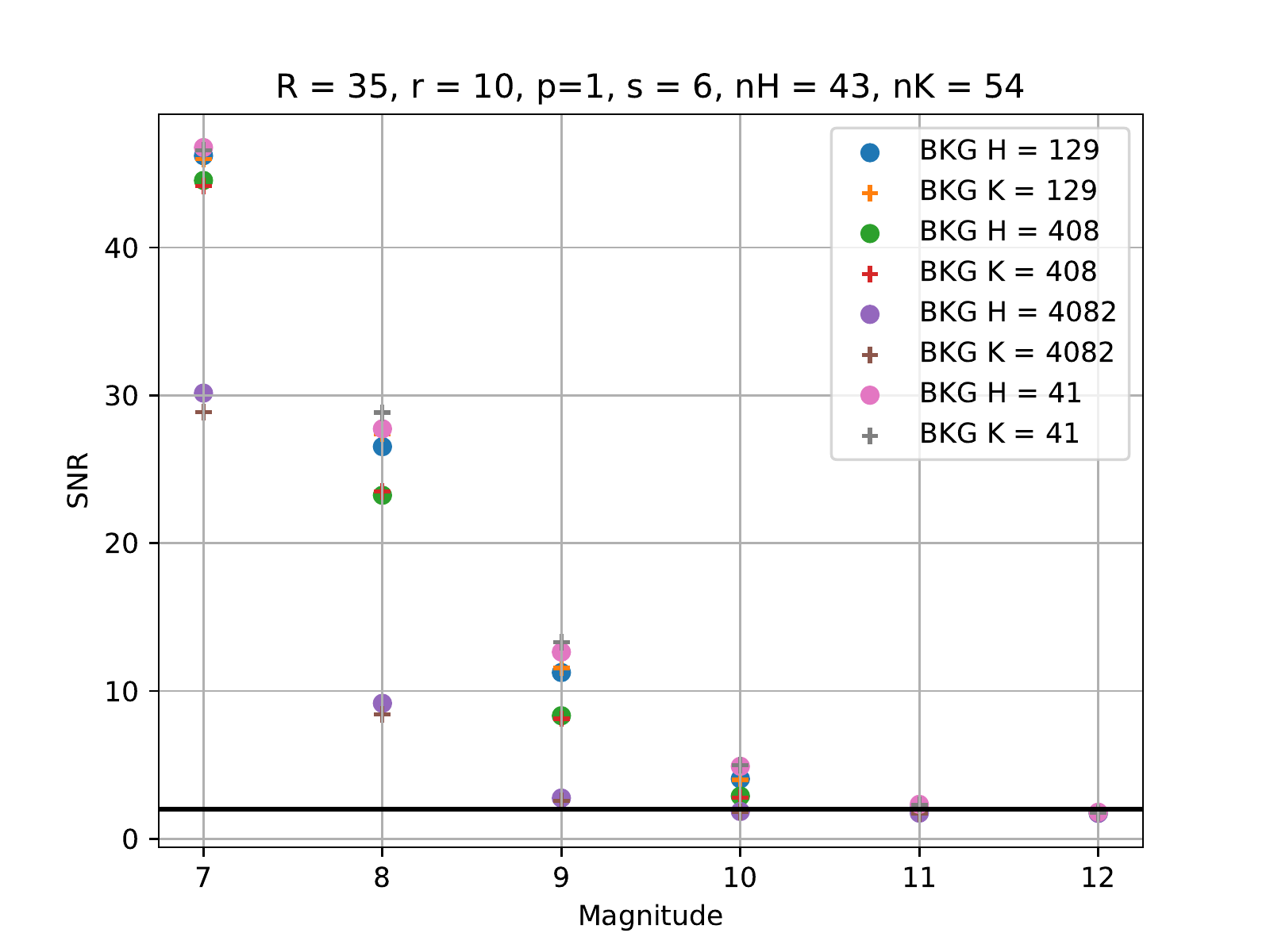}
\includegraphics[width=0.5\textwidth]{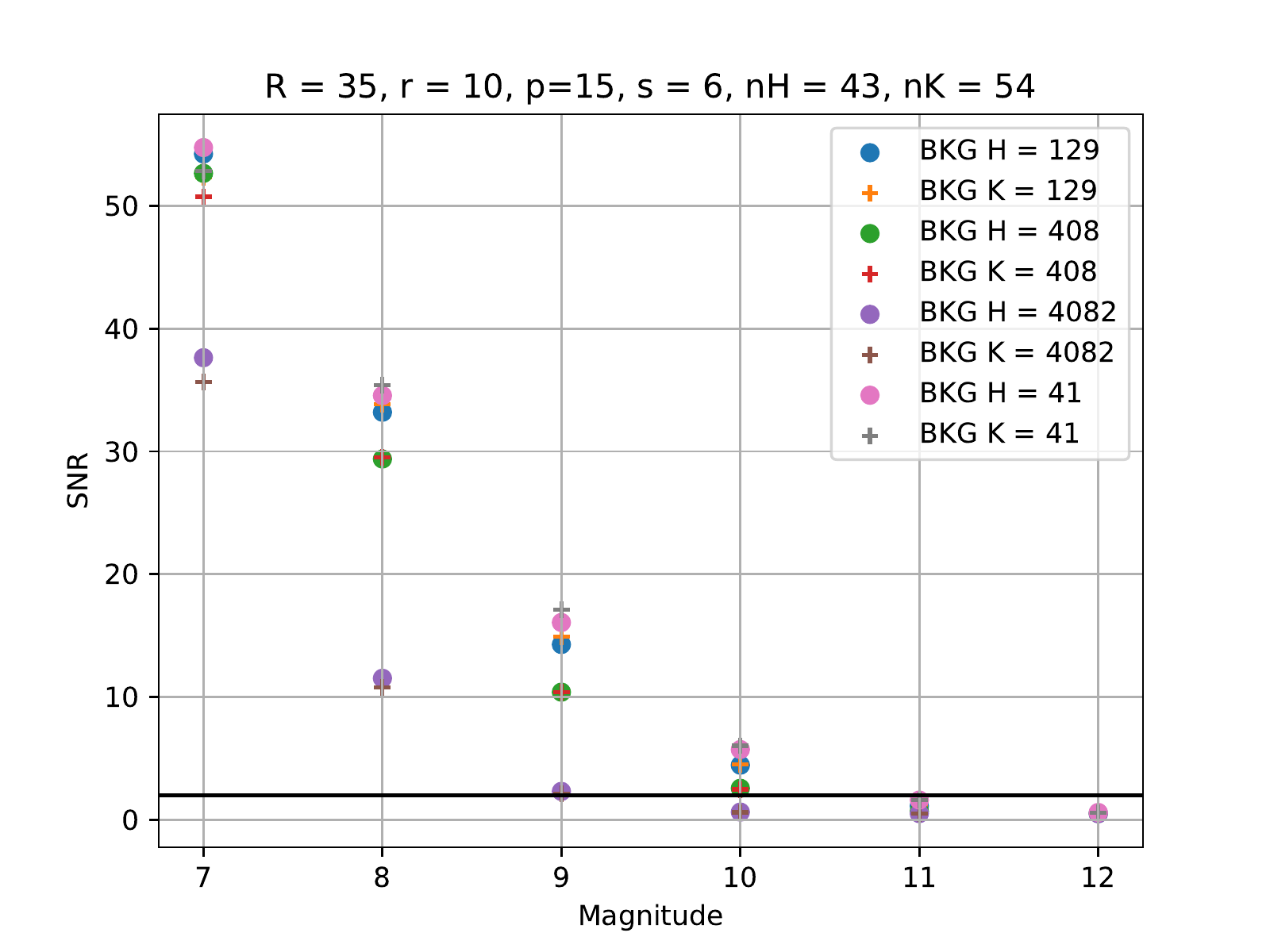}}
\caption{SNR as a function of the star magnitude for different values of coherent adding of frames in the power spectrum. The parameters used on these plots are $r_0$ = 10 cm, R = 35, n = 43, s = 6. The left plot is for p = 1, and the right is p = 15 frames added incoherently.}\label{fig:p}
\vspace*{0.5cm}
\end{figure}

In Fig.~\ref{fig:p}, we display the SNR as a function of the magnitude. The dots are for the H-band simulations, the crosses are for the K-band simulations. The different colors represent different levels of background with the values in legends shown in e/s/pixel. The black line is showing the SNR = 2, the limit for which we should be able to track the fringes (phase tracking). The different plots are for different numbers of power spectra coherently added (p = 1 and 15 frames). Note that the trend on the plot is similar for the other values of p not shown here.

We can notice that for lower magnitude (7, 8, and 9) the SNR for a same background level is improving with increasing p. However the tendency is reverted for higher magnitude (10, 11, 12), the SNR decreasing when increasing p.

\begin{figure}[h!]
\makebox[\textwidth]{\includegraphics[width=0.7\textwidth]{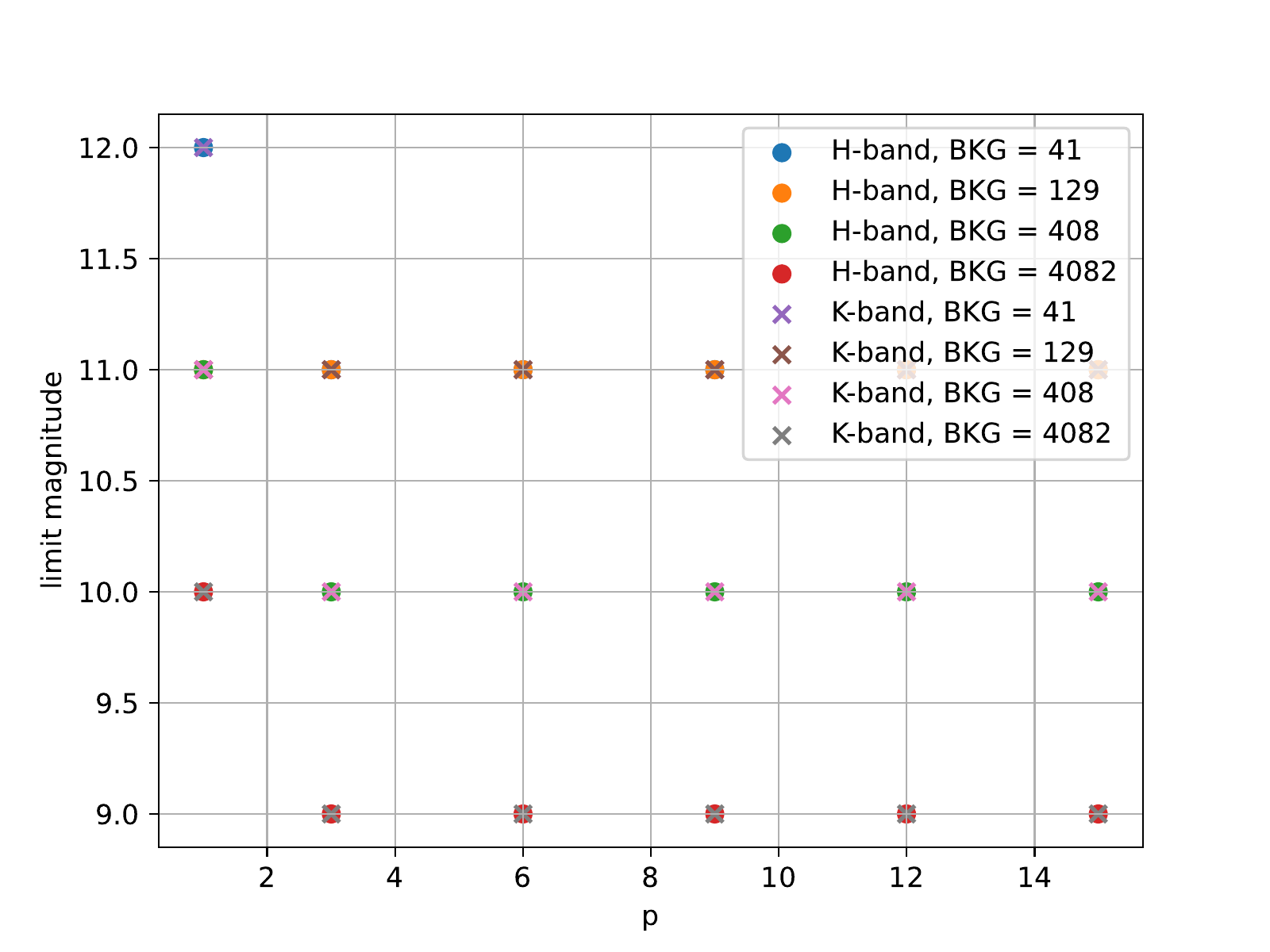}}
\caption{Limiting magnitude as a function of p, for different background levels (colors) and the different spectral bands (dots: H-band, crosses: K-band)}\label{fig:p_mag}
\vspace*{0.5cm}
\end{figure}

In Fig.~\ref{fig:p_mag}, we show the limiting magnitude as a function of p, for different backgrounds level and the two different spectral bands.
We can first see that the limiting magnitude is the same for both spectral bands for each different set of parameters. We can also see that the limiting magnitude decreases when p increases, which is unexpected. We can also notice that the limiting magnitude does not vary for $\text{p} > 1$. It appears then that for faint stars, the best value is $\text{p} = 1$. For $\text{p} = 1$, the best limiting magnitude is 12, but only for the lowest background value. The two intermediate background levels have a limiting magnitude of 11. The highest background has a limiting magnitude of 10. For $\text{p} > 1$, the two lowest background levels have the same limiting magnitude of 11. For BKG = 408e/s/frame, the limiting magnitude is then 10. Finally, for the highest background level, the limiting magnitude is 9.

\subsection{ Effect of the seeing ($r_0$) on the SNR}

Changing the $r_0$ parameters allows us to see how the magnitude limit will be impacted by the different seeing conditions. The implementation of the AO system on all telescopes should improve the bad and normal seeing conditions to mostly good seeing conditions.

\begin{figure}[h!]
\makebox[\textwidth]{
\includegraphics[width=0.5\textwidth]{figures/r10_R35_n43_p1_s6.pdf}
\includegraphics[width=0.5\textwidth]{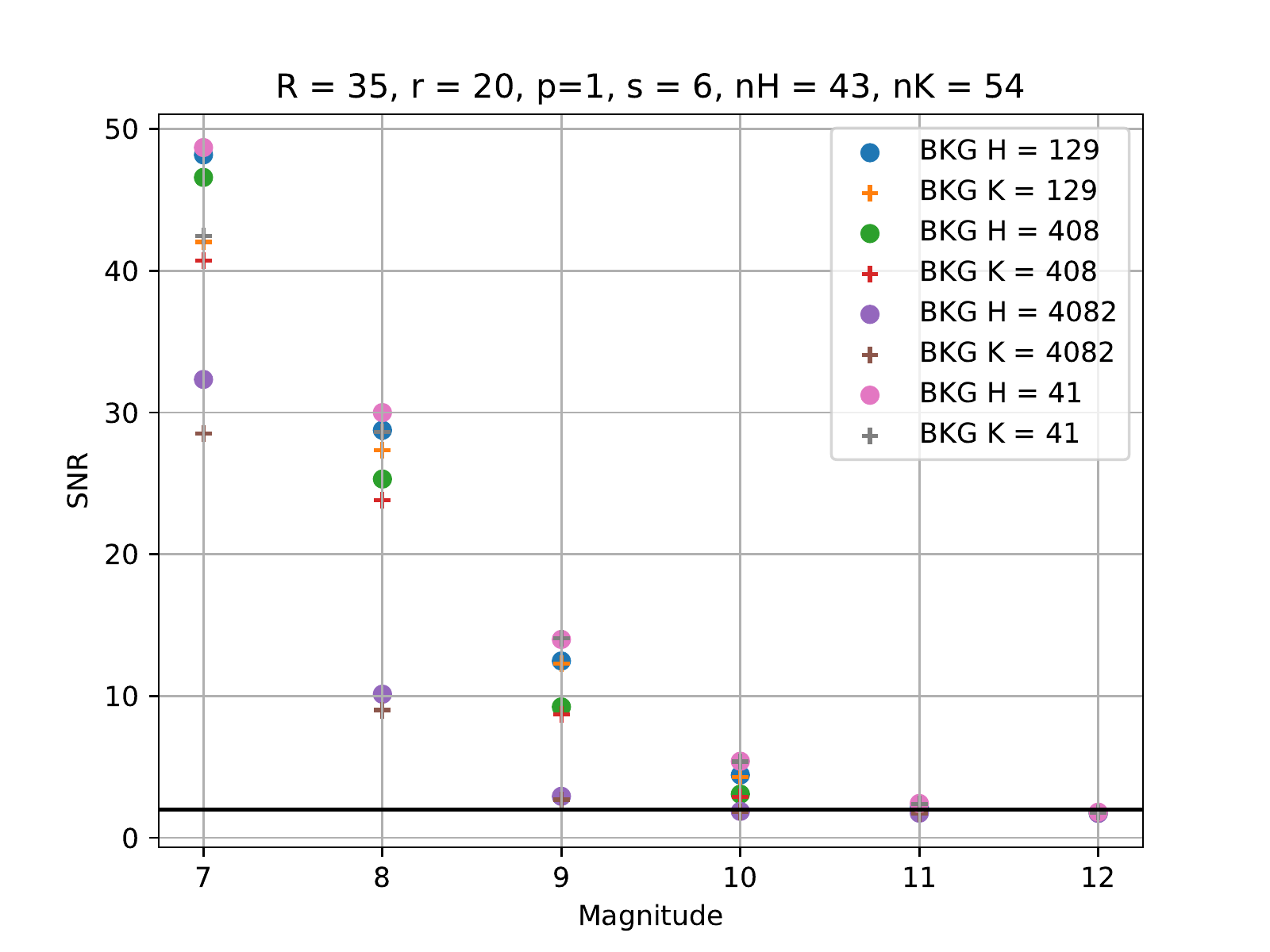}}
\caption{SNR as a function of the star magnitude for different values of $r_0$. The parameters used on these plots are R = 35, n = 43 pixels, p = 1, and s = 6 frames added incoherently. The left plot is for $r_0$ = 10 cm, and the left is $r_0$ = 20 cm.}\label{fig:r0}
\vspace*{0.5cm}
\end{figure}

In Fig.~\ref{fig:r0}, we display the SNR as a function of the magnitude; the dots are for the H-band simulations and the crosses are for the K-band simulations. The different colors represent different levels of background with the values in the legends given in e/s/pixel. The black line is showing the SNR=2 case, the limit for which we should be able to track the fringes (phase tracking). The different plots are for different values of $r_0$. Note that the trend is similar for $r_0$ = 5 cm, not shown here.

We can see in Fig.~\ref{fig:r0} that the $r_0$ does not have much effect in the K-band, which is expected as the K-band turbulence is more stable than the turbulence in the H-band.
In the H-band, we see an improvement with the $r_0$ parameter increasing. But its effect is mostly noticeable at lower magnitudes. The effect at high magnitude (fainter targets) is too small to have an effect on the limit observable magnitude. 
We can see that with normal atmospheric conditions, most of the background conditions result in a limiting magnitude of 10, while a magnitude of 11 needs better background and is at the limit of being observable.

\begin{figure}[h!]
\makebox[\textwidth]{\includegraphics[width=0.7\textwidth]{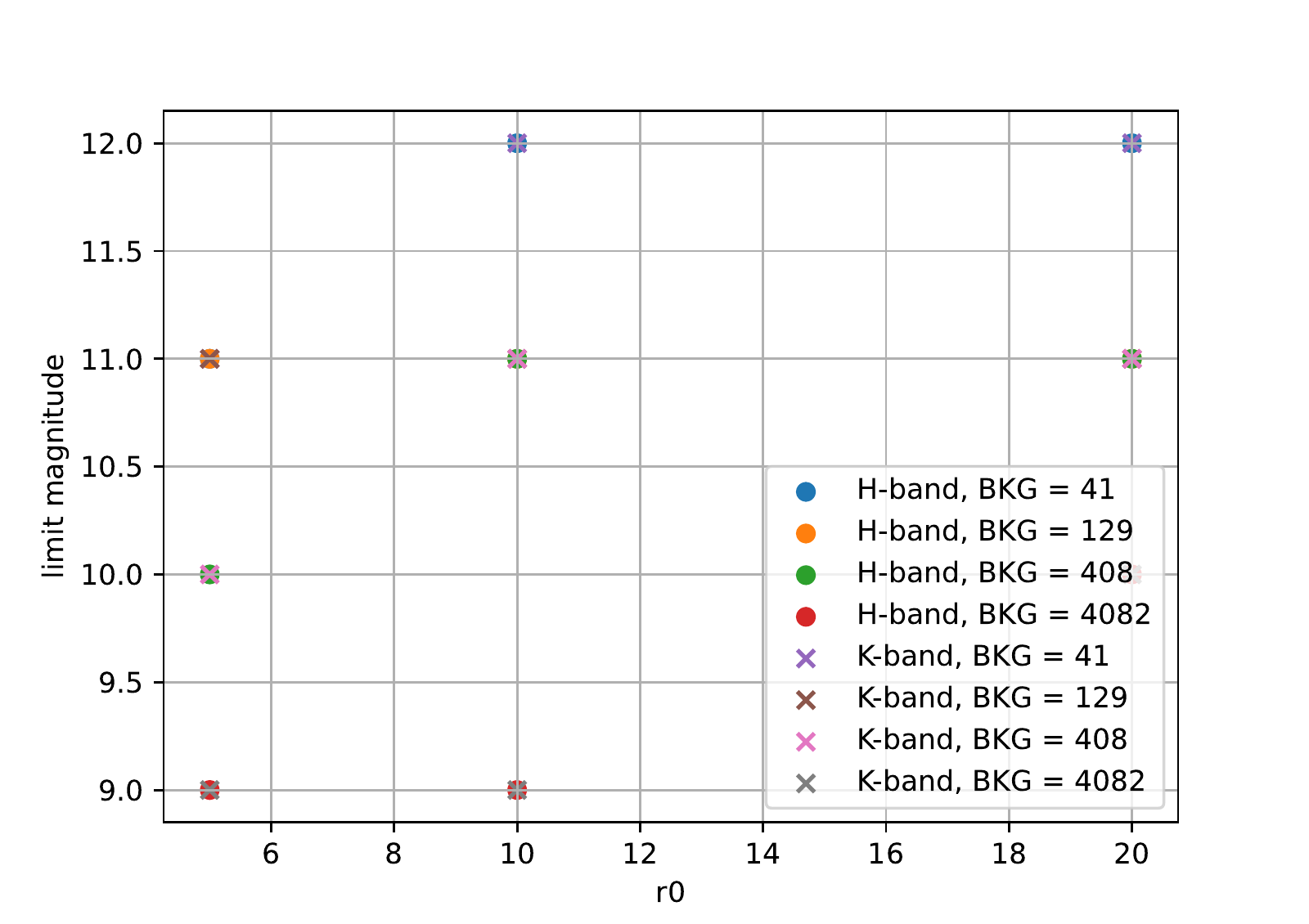}}
\caption{Limiting magnitude as a function of $r_0$, for different background levels (colors) and the different spectral bands (dots: H-band, crosses: K-band)}\label{fig:r0_mag}
\vspace*{0.5cm}
\end{figure}

In Fig.~\ref{fig:r0_mag}, we show the be limiting magnitude as a function of p, for different backgrounds level and the two different spectral bands.
We can see that the limiting magnitude is increasing with the $r_0$, which means that we can observe fainter stars with better atmospheric conditions, which is expected. We also see that for bad conditions ($r_0$ = 5) the limiting magnitude for both the lower background levels is 11. For BKG = 408, the limiting magnitude is 10. For the highest background level, the limiting magnitude is 9. We reach a limiting magnitude of 12 in normal good atmospheric conditions (respectively $r_0$ = 10 and $r_0$ = 20) for the lowest background level. For the two intermediate background levels, we reach a limiting magnitude of 11 in the same conditions. For the lowest background level, we need to have good atmospheric conditions to reach the limiting magnitude of 10.

\subsection{Effect of the number of pixels in the fringe dimension (n) on the SNR}

This parameter allows us to see the effect of the number of pixels we use to probe the fringes in the fringe direction. It is helpful to inspect the effect of changing the value of the number of pixels we use to probe a single fringe, but it is actually not useful to vary the number of pixels needed to probe different wavelengths (fixed for now to 2.5 pixel/fringe). The more pixels we use to probe the fringes, the more accuracy we obtain to compute the contrast of the fringes, but more light is needed to be spread over several pixels, losing SNR.

\noindent
\begin{figure}[ht!]
\makebox[\textwidth]{\includegraphics[width=0.5\textwidth]{figures/r10_R35_n43_p1_s6.pdf}
\includegraphics[width=0.5\textwidth]{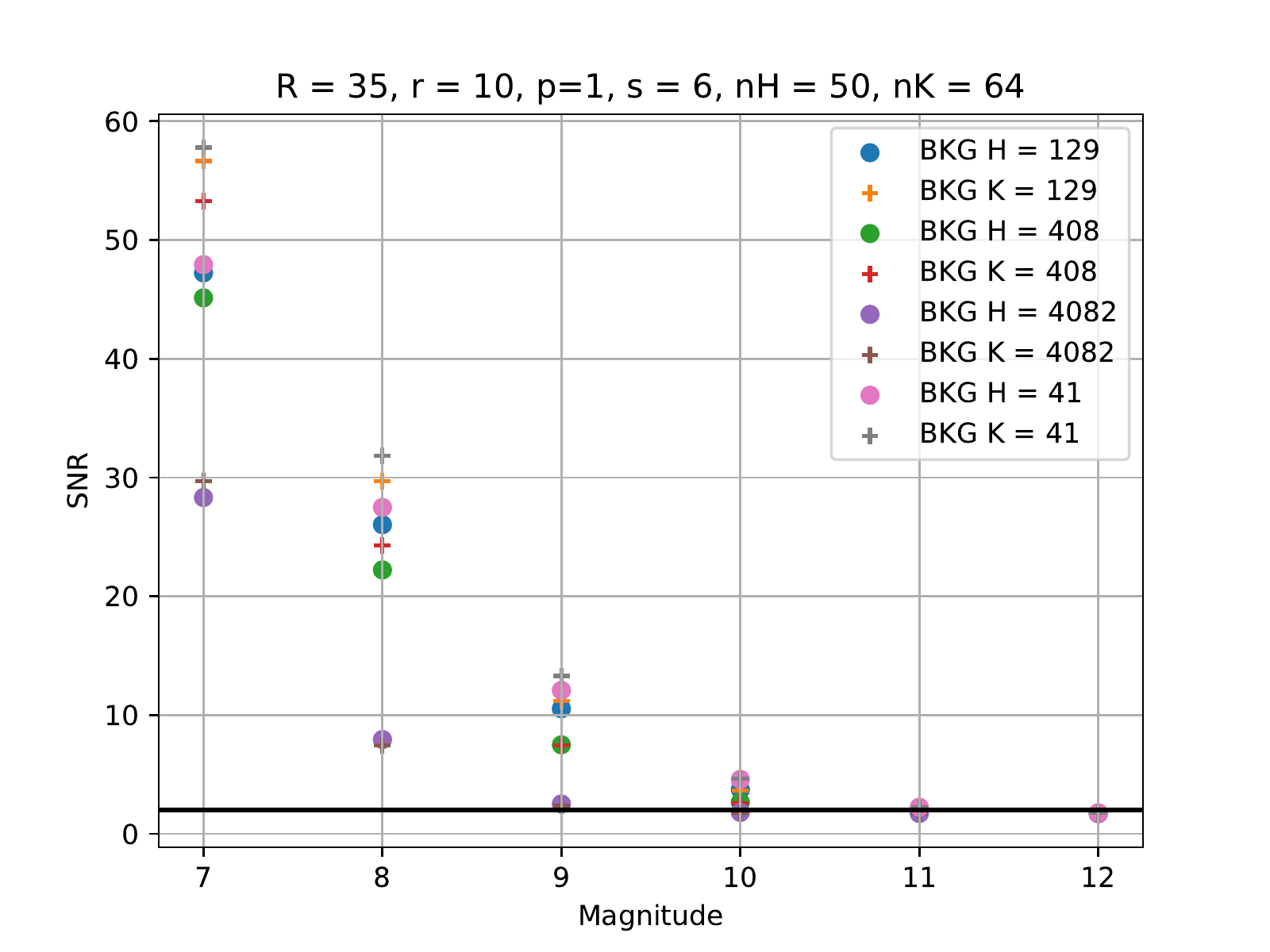}}
\caption{SNR as a function of the star magnitude for different values of n. The parameters used on these plots are $r_0$ = 10 cm, R = 35, p = 1, and s = 6 frames added incoherently. The left plot is for n = 43 pixels, and the right is n = 50 pixels in the H-band, which gives us n = 54, and 64 respectively in the K-band.}\label{fig:n}
\vspace*{0.5cm}
\end{figure}

In Fig.~\ref{fig:n}, we display the SNR as a function of the magnitude.  Dots portray the H-band simulations while the crosses are for the K-band simulations. The different colors represent different levels of background with the values in the legends being in e/s/pixel. The black line shows the SNR=2, the limit for which we should be able to track the fringes (phase tracking). The different plots are for different numbers of pixels that are probing the fringes. Note that the trend is the same for the values of n not shown here.

In Fig.~\ref{fig:n}, the number of pixels we use in the fringe direction seems not to affect the SNR in the H-band for bright stars. However, the more pixels we use to probe the K-band, the higher the SNR for those same bright stars. On the other hand, for faint stars, the H-band seems to be more affected than the K-band, with the SNR going slightly lower for higher numbers of pixels in the fringe direction.

\subsection{Summary on the study of the expected performances of SILMARIL}

With ideal conditions and the lower background level, we should be able to observe stars with a magnitude up to 12, by incoherently adding 12 frames. In most conditions, by co-adding only 3 frames, we should be able to observe magnitudes of 10 for most of the background conditions. This limiting magnitude corresponds to an improvement of about 2 to 4 magnitudes compared to CLIMB, which is in agreement with the estimates we computed without the simulations. The best parameters in normal to good conditions would be p = 1 and s = 12 for faint stars.

\section{Real-time software architecture and data reduction pipeline}\label{sec:soft}

For minimum software development efforts and to reuse much of the working code, SILMARIL adopts the MIRC-X software \cite{anugu2020} for both the real-time software and the data reduction pipeline as there are several similarities between SILMARIL with MIRC-X.  
\begin{itemize}
    \item \textbf{Detector data acquisition:} The MIRC-X software directly (as-it-is) can be used for the detector data acquisition and critical instrument security and health monitoring. Both these instruments use the (i) same C-RED One camera, (ii) data acquisition hardware (Matrox frame grabber and fiber camera-link extender system), and (iii) same ion pump vacuum and safety system.  The data acquisition software runs on a similar computer and grabs images from the C-RED One camera using camera-link cables with a dedicated Matrox frame grabber in real-time with a low latency Linux operating system \cite{anugu2020}. 
    \item \textbf{Fringe Acquisition and Group-delay Tracking:} MIRC-X uses six beams, while SILMARIL uses three-beams only, optimized for sensitivity. We reuse the MIRC-X group delay fringe tracking engine \cite{anugu2020}.
    \item \textbf{Data reduction pipeline:} We plan to make similar data acquisition sequences.  For the data reduction, we plan to adopt the MIRC-X data reduction pipeline \cite{anugu2020} written in python for SILMARIL. The only major difference is the change to three beams.
\end{itemize}

\section{Science Cases}\label{sec:sci}

The gain in sensitivity SILMARIL will open numerous new science cases not achievable yet and will push the science cases already on-going at the Array. Here we present a few science cases for which SILMARIL will be used.

\subsection{AGNs}
Over the last few years, the study of AGNs has been going through a transformation. In the standard picture, which has been around for more than 30 years\cite{1985ApJ...297..621A}, an obscuring torus is invoked and thought to surround the accreting supermassive black hole at the center. The physical origin of this torus is not clear, but it is assumed to be more or less static, and importantly, it unifies the two major AGN categories: those with a face-on, polar, direct view of the nucleus, called Type 1, and those with an edge-on, equatorial, hidden nuclear view, called Type 2. Recent mid-IR interferometry has shown that a major part of the mid-IR emission, believed to be coming from the outer warm ($\sim$300K) part of this putative dusty torus, has a polar-elongated morphology, rather than the expected equatorially elongated structure\cite{2012ApJ...755..149H,2013ApJ...771...87H,2014A&A...565A..71L}. In addition, this polar-elongated dusty gas is in fact UV-optically-thick, since the measured IR emissivity is a few tenths and consistent with directly illuminated UV-optically-thick gas. Furthermore, at the same spatial scale, ALMA finds a polar outflow~\cite{2016ApJ...829L...7G}, likely to be an inward extension from the 10-100pc scale bipolar outflow directly resolved by HST\cite{2002AAS...200.0510C}.

The new scenario is that the torus is an obscuring and outflowing structure, probably being driven by the radiation pressure on dust grains from the anisotropic, polar-strong radiation field of the central accretion disk. This dusty wind is likely giving strong feedback to the host galaxy, resulting in the correlation between the bulge and central black hole masses in galaxies\cite{2000ApJ...539L...9F}. 

If this picture is correct, the wind must be launched at the innermost radius where the dust grains barely survive, that is in the dust sublimation region. Does this region show a substantial scale height, leading to a polar-elongated morphology at this innermost radius as well? Since the dust sublimation temperature is supposed to be $\sim$1500 K, the region must be the brightest at $\sim$2 $\mu$m, i.e., in the K-band. Up to now, the region has only been marginally resolved by interferometers at $\sim$100 m baselines\cite{2011A&A...536A..78K,2012A&A...541L...9W,2020A&A...635A..92G,2021ApJ...912...96L}. 

With the CHARA Array’s long baselines of up to 330 m, we will be able to map the elongation of the structure at a sub-mas or 0.1 pc scale in the near-IR for the first time. That will be a fundamental test of the new picture, and a major breakthrough if any elongation is detected at this critical scale. There is a closely related, but very different aspect in this nuclear region study. A supermassive black hole is now believed to reside in essentially every galaxy. Naturally, after collisions and mergers of these galaxies, a binary black hole would form at the center. Theoretical studies with numerical simulations have suggested that the binary separation quickly shortens by ejecting the surrounding stars, but could stall at the 0.1 pc scale known as the "final parsec problem"\cite{2005LRR.....8....8M}. This is consistent with recent Pulsar Timing Array constraints, where low-frequency gravitational waves, expected from supermassive black hole mergers, are not being detected\cite{2015Sci...349.1522S}. A big question is: are these massive binaries really there? We simply do not have direct, spatially resolved constraints at this 0.1 pc scale. However, the CHARA Array can drastically change this situation. We might identify a binary structure directly at the center of an AGN for the first time.

\begin{figure}[h]
\makebox[\textwidth]{\includegraphics[width=0.5\textwidth]{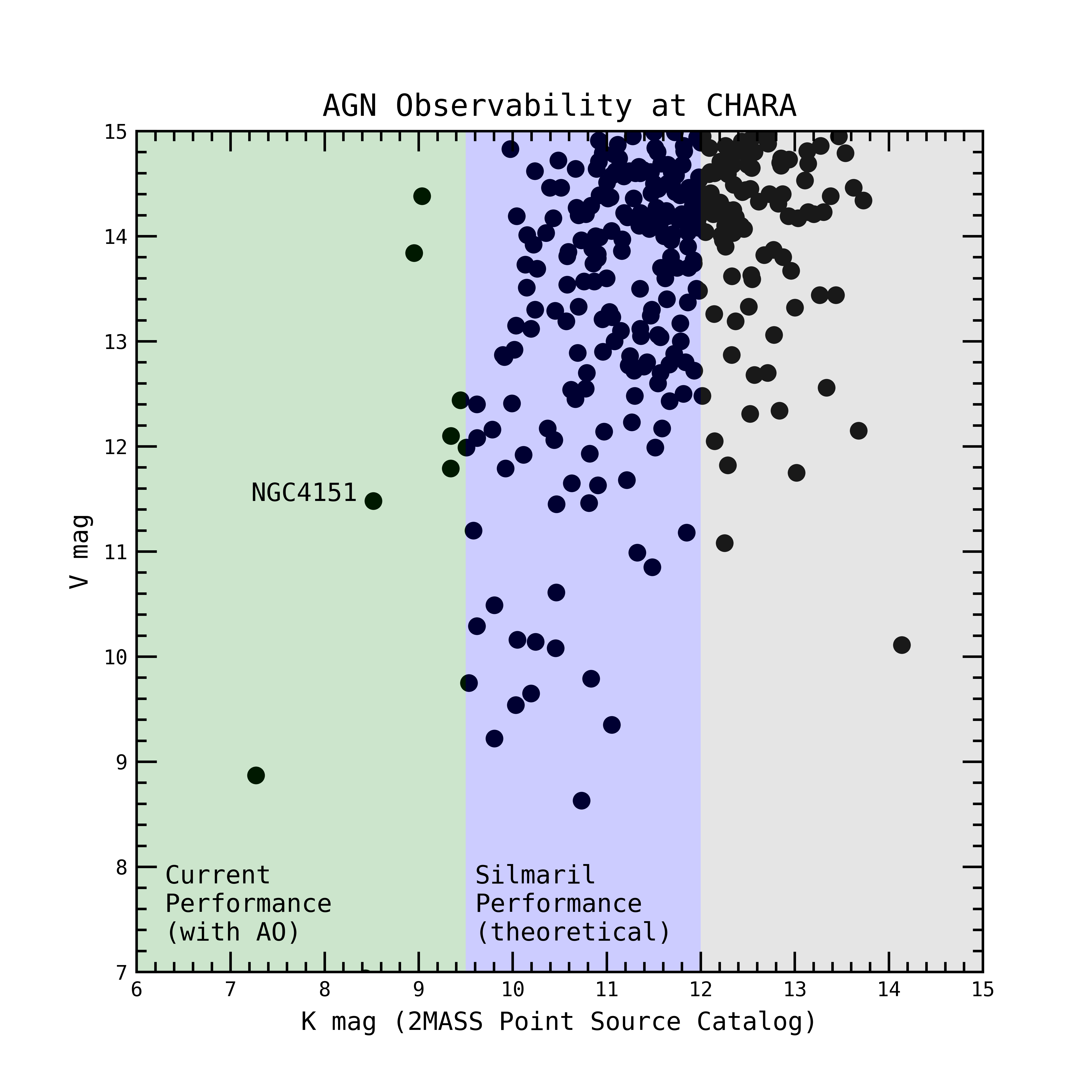}}
\caption{Magnitude in V-band as a function of magnitude of K-band for AGNs. In green: already observable at CHARA. In blue: observable with SILMARIL
}\label{fig:AGNimp}
\end{figure}

We have already demonstrated that we can detect fringes on the brightest Type 1 AGN NGC4151 and resolve its geometry on long 250 m baselines using CLASSIC (Kishimoto, Anderson, ten Brummelaar et al.~submitted). Using the blind observing technique and a sensitivity improvement of at least $\sim$1.5 magnitudes should allow us to observe a number of AGNs and help to resolve several of these issues (Fig.~\ref{fig:AGNimp}).

\subsection{Angular Diameter of Stars}

Angular diameter measurements are used to determine empirical surface brightness relations for predicting radii based on photometric colors~\cite{2014AJ....147...47B,2018MNRAS.473.3608A, 2021A&A...652A..26S}. Combined with a precise Gaia parallax\cite{GAIADR2} and a measurement of the bolometric flux, the angular diameter can be used to derive the physical radius and effective temperature of a star. At the low-mass end of the main sequence, predictions from evolutionary models tend to overestimate the temperatures of stars by 3\% and underestimate the radii of stars by 5\%~\cite{2012ApJ...757..112B}. Comparing radii and temperatures with evolutionary models provides a way to measure the ages of nearby moving groups~\cite{2011ApJ...743..158S,2015ApJ...813...58J,2018ApJ...858...71S,2022AJ....164...34M} and exoplanet host stars~\cite{2009ApJ...701..154B}. Moreover, these fundamental parameters are used to refine the location of the habitable zone around exoplanet host stars~\cite{2011ApJ...729L..26V,2014MNRAS.438.2413V,2015ApJ...800..115T} and infer the radius of transiting exoplanets based on the stellar diameter and eclipse timing~\cite{2011ApJ...740...49V}. With the successful launch of the TESS mission, the number of exoplanet hosts whose radii can be measured directly by CHARA will increase dramatically; there are  $\sim$6500 stars in the TESS input catalog within the resolution and sensitivity limits of CLASSIC. Targets with solar-like oscillations can be used to calibrate asteroseismic scaling relations for measuring the masses and radii of stars~\cite{2012ApJ...760...32H}. The improvement in the sensitivity limit will not impact the number of TESS targets accessible, because the stars with H $>$ 7 are too small to resolve even with the CHARA Array. However, improvements in precision and efficiency will have a dramatic effect on the quality and number of diameters measured.

\subsection{Disks around Young Stars}

The CHARA Array is unmatched for studying circumstellar disks around young stellar objects (YSOs) because it possesses the unique combination of long baselines ($>$200 m) and sensitive instrumentation in the infrared. CHARA was the first interferometer to confirm the detection of hot gas inside the dust sublimation radius~\cite{2008PhDT........11T,2008SPIE.7013E..0UT}. Most young stellar objects are faint (H, K $>$ 7) and require acquiring low-contrast fringes. With the increased sensitivity of ESO’s Very Large Telescope Interferometer (VLTI) a large number of YSO disks have been observed in recent surveys conducted by VLTI/PIONIER~\cite{2017A&A...599A..85L} and VLTI/GRAVITY \cite{2019A&A...632A..53G}. 

VLTI instruments are limited by the spatial resolution offered by its maximum baseline of 130m. With nearly triple the resolution of VLTI, CHARA provides opportunities to study the inner structure of disks around YSOs. With improved sensitivity brought by SILMARIL, a larger survey of the inner AU of protoplanetary disks will reveal the importance of gas emission and stellar winds, and shed light on puffed-up inner walls. It should also be possible to detect accreting protoplanets embedded in the disk~\cite{2015Natur.527..342S}. An improvement in sensitivity of 1-2 magnitudes with SILMARIL will increase the sample size by a factor of 2-3.

\subsection{Winds from Massive Stars}

The extreme luminosity of massive stars drives stellar winds that may carry away a large fraction of a star’s mass over its lifetime. The highest mass loss rates are found among the Luminous Blue Variables (LBVs), and Richardson et al. (2013)~\cite{2013ApJ...769..118R} used the CHARA Array to resolve the wind outflow in the nearest LBV, P Cygni. Two other LBVs will be accessible with greater sensitivity, HD 168607 and V446 Scuti, as well as dozens of distant Wolf-Rayet stars. Many of these objects experienced episodes of very large mass loss or systemic mass loss in a binary system, such as the pinwheel outflow in WR~104~\cite{2008ApJ...675..698T}. CHARA observations will help map these tracers of mass loss processes.

\subsection{Interacting Binary Systems}

The first instance of Roche lobe overflow in massive binaries may result in large-scale mass loss from the system and the creation of a circumbinary disk such as the huge torus surrounding RY Scuti~\cite{2011MNRAS.418.1959S}. The W Serpentis and FS CMa binaries probably represent systems in this intense stage of mass loss, and several dozen targets will be accessible with a more sensitive detector. Binaries that survive the transformation of the first evolving star into a neutron star or black hole remnant will form a Massive X-Ray Binary system, and some dozen MXRBs will be accessible with better sensitivity to explore their mass loss processes. For example, the GRAVITY Collaboration at VLTI recently resolved some of the inner structure of the MXRB SS 433, and CHARA observations would probe deep into the central engine that launches the relativistic jets~\cite{2017A&A...602L..11G}. 

Massive stars are often found in multiple systems, and any wide companions of MXRBs discovered by CHARA would help establish the reflex orbits of the visible mass donor stars in these systems~\cite{2014PASA...31...16M}. CHARA observations would also probe mass transfer processes in wide symbiotic binaries~\cite{2007BaltA..16..104F} and end-of-life loss into circumbinary disks surrounding post-AGB stars~\cite{2018arXiv180900871V}.
The gain in sensitivity with SILMARIL would open these targets to investigation with CHARA.

\subsection{Young Binary Systems}

\begin{figure}[h]
\makebox[\textwidth]{\includegraphics[width=0.5\textwidth]{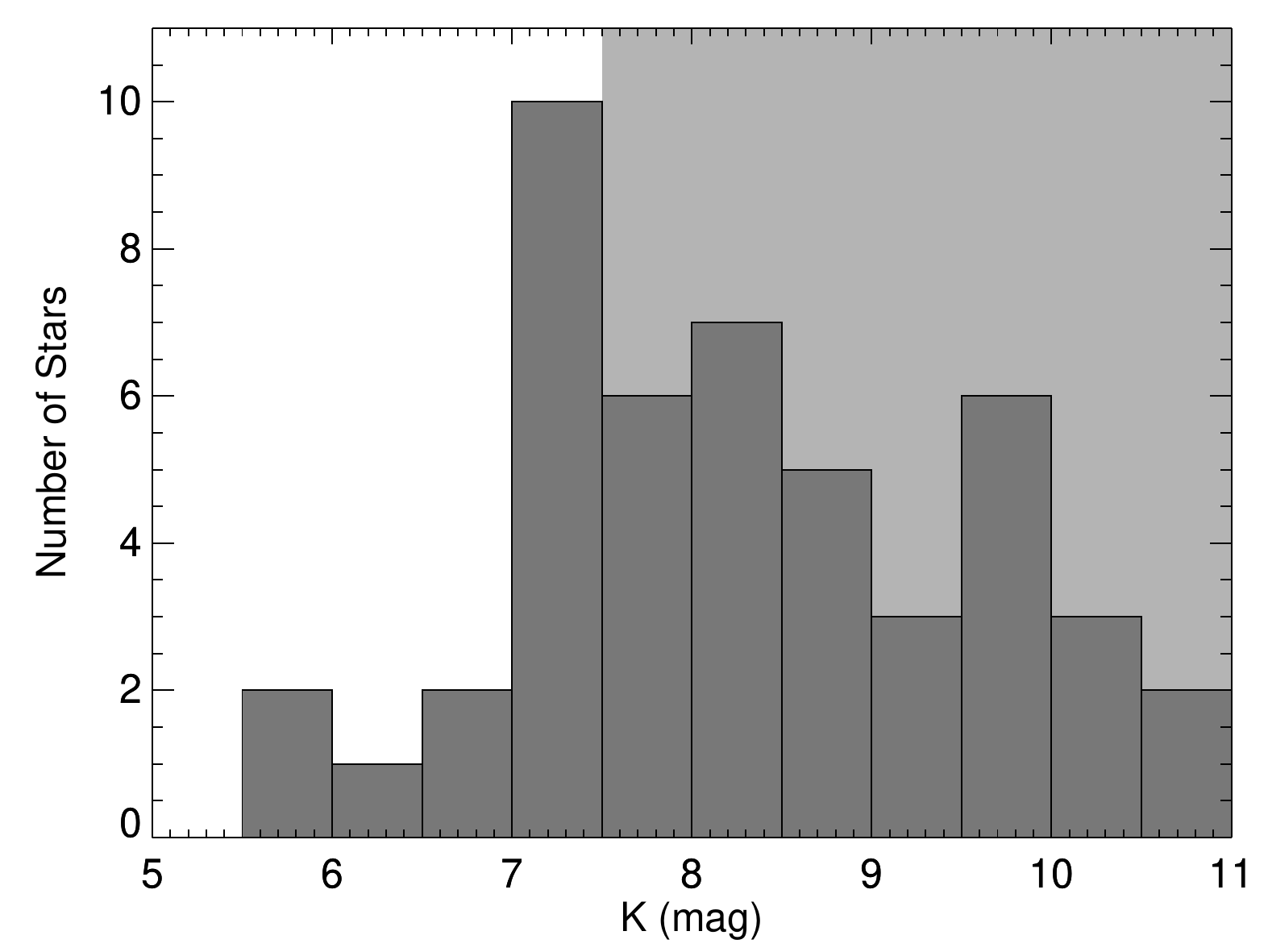}}
\caption{Histogram of young spectroscopic binaries in the nearby star forming regions accessible by the CHARA Array.
}\label{fig:specbin}
\end{figure}

Spatially resolving the orbits of double-lined spectroscopic binaries yields the component masses and distance to the system. Many spectroscopic binaries in nearby star forming regions (Taurus, Orion, Ophiuchus) with ages $<$ 10Myr are just beyond the reach of the current sensitivity limits at the CHARA Array. Figure~\ref{fig:specbin} shows a histogram of known spectroscopic binaries in nearby star forming regions accessible to the CHARA Array. At the current magnitude limit of the MIRC-X combiner in excellent seeing conditions (H $<$ 7.5), only the brightest binaries in most of these regions are resolvable. To extend to lower mass members, where evolutionary tracks are most discrepant at young ages\cite{2017ApJ...841...95S}, an improvement in sensitivity is required. Dynamical masses can be used to calibrate the models of stellar evolution at young ages. Moreover, for a given set of tracks, dynamical masses across a range of spectral types provide a strong constraint for age dating of clusters, which impacts our understanding of the history and chronology of planet formation in these regions. 

\subsection{Transient Events}

In the era of large time-domain surveys like LSST and the Zwicky Transient Survey, the CHARA Array can be used to follow-up bright transient events to measure the spatial extent of the source. This includes the ability to resolve the angular expansion and development of asymmetries during the early stages of nova explosions~\cite{2014Natur.515..234S}. Another promising application would be to measure the image size and centroid displacement in gravitational microlensing events; this would break the degeneracy between the separation and mass ratio of a planet, brown dwarf, or black hole companion relative to the host lens star\cite{2016MNRAS.458.2074C,2019ApJ...871...70D}.

\section{Conclusions}\label{sec:concl}
Thanks to the new e-APD technology, optical long baseline interferometry is able to attain better sensitivity. With a design that uses both proven concepts, such as a minimum number of optics, and new concepts, such as the edge filter, we can push for even more sensitivity with SILMARIL. The expected performances once achieved with an external dewar should bring the limiting magnitude to 11 both in H- and K-bands in average atmospheric conditions, and to 12 in the best atmospheric conditions. This will open a number of new science cases not achievable yet at the CHARA Array and extend the range of already ongoing science.

The implementation of SILMARIL is ongoing, with a first test on sky scheduled in September 2022, with an engineering camera. The C-RED One camera is expected to be delivered around the end of the year 2022. Once the efficiency of the instrument will be proved on sky, we will work on the implementation of the external dewar that will push its performance further, and on the implementation of the second set of 3 beams to make a use of the 6 telescopes of the CHARA Array.

\appendix

\acknowledgments 
The SILMARIL project at the CHARA Array is supported by the National Science Foundation under Grant No.  
AST-1909858.  
Institutional support has been provided from the GSU College of Arts and Sciences 
and the GSU Office of the Vice President for Research and Economic Development.

\bibliography{report} 
\bibliographystyle{spiebib} 

\end{document}